# Quantum coupled-wave theory of price formation in financial markets: price measurement, dynamics and ergodicity

**Jack Sarkissian**

Managing Director, Algostox Trading

New York, NY

Abstract

We explore nature of price formation in financial markets and develop a theory of bid and ask price dynamics in which the two prices form due to quantum-chaotic interaction between buy and sell orders. In this model bid and ask prices are represented by eigenvalues of a 2x2 price operator corresponding to "bid" and "ask" eigenstates, while randomness of price operator results in price fluctuations that destroy oscillatory effects. We show that this theory adequately captures behavior of bid-ask spread and allows to model bid and ask price dynamics in a coordinated way. We also discuss ergodicity properties of price formation and show how directional price movement occurs due to ergodicity violation in a quantum process instead of the commonly believed forces acting on price. This theory has wide range of applications such as trade execution modeling, large order pricing and risk valuation for illiquid securities.

## 1. Stochastic calculus vs. financial markets

Modern quantitative finance is built around stochastic calculus treating price movements as random walk [1-3]. In doing so it implicitly relies on certain assumptions, that are supposed to ensure similarity between the two processes. Over the decades, countless models have been produced based on this similarity. However, in practice these models are almost always used without verifying the validity of the underlying assumptions. Such careless application of models can easily lead to lack of control, position mismanagement and financial losses [4-6]. Let us therefore examine, what criteria ensure similarity between price movements and random walk.

Stochastic calculus assumes that security price is always available. Indeed, for any point in time suffice it to propagate the stochastic process up to that point and the price result is guaranteed. The reality is, however, different. In institutional finance many assets do not have a current price. Many assets are only traded a few times a day, many haven't been traded for months, and many have never been traded before, Fig. 1. In the meantime, it is these "illiquid" securities that make up the majority of institutional trading. The "price availability" assumption may be suitable only for well traded, liquid securities, such as for example the blue-chip stocks. And even then, frequent transactions do not guarantee price availability, because price availability for 200 shares does not automatically imply price availability for an institutional level million-share block.

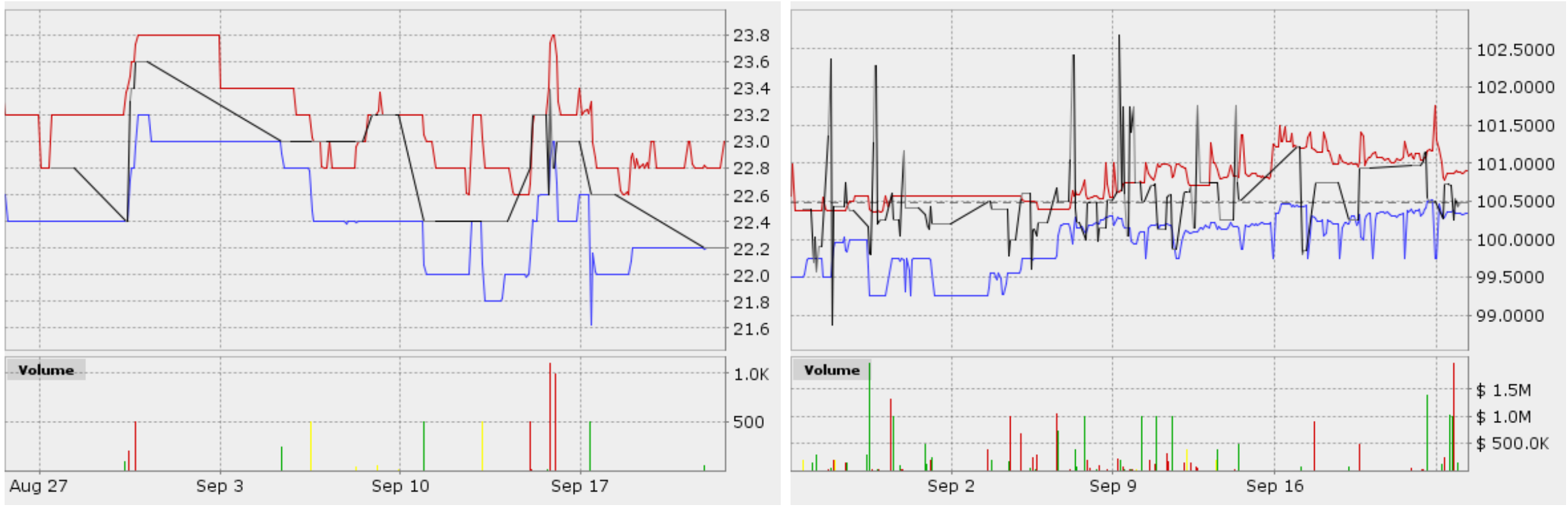

Fig. 1. Typical institutional class securities. Red line represents asking price, blue line represents buying price, black line represents last trade price. Left: FRIWO AG stock (CEA @ XETRA). Traded on exchange a few times a month and in very small quantities. Right: MPLX 6 7/8 bond, July 31 2019 issue. Even though the security is traded every day, it is illiquid for an order size of over $1 mln.

Another significant simplification of stochastic calculus is its treatment of price as a single number. The idea that buying and selling prices are in fact different does not need elaboration - it has been common knowledge for ages. This difference is especially visible in institutional trading where transaction size is large. For example, it is very common in fixed income markets to have the buying and selling prices set apart by 50 bp[1] or 250 bp. The reality is even more colorful, since often securities would have no buying price (no immediate buyer) or no selling price (no immediate seller), and it is not unusual to have neither. The assumption that buying and selling prices are equal rests on implication that (a) securities are liquid (well-traded) and (b) order size is small. For such securities the buy-sell price difference is small compared to price itself and can be neglected: $\frac{s_{sell}-s_{buy}}{0.5(s_{sell}+s_{buy})} \ll 1$. Even then, some trading desks and entire firms, called market makers, live off this tiny difference by turning securities over very quickly and capturing this difference many times.

[1] Basis point (bp) is 1/100-th of 1%, common measure of interest and spread in finance.

The only time when it is valid to say that price is a single number is at the time of transaction. Transactions occur when buy and sell prices match, thus reducing price to a single number. That case is very special and deserves detailed consideration, which will be given in the next section.

Stochastic calculus has made its way into finance mainly by providing a seemingly solid framework for modeling financial instruments. However, as we see, its area of applicability is in fact limited to small orders in liquid, well-traded securities. These conditions are typical for small retail accounts, such as someone's individual brokerage account or a retirement account. They are generally inapplicable for institutional level operations. Financial institutions cannot rely on analytics for billion-dollar portfolios that is based on prices relevant for only 200 shares and worth only tens of thousands of dollars. It is impossible to rely on risk reports with a 10-day time horizon when it takes 3 months to liquidate the portfolio at a steep discount. Lastly, it's against common sense to use the same price valuation for buying and selling orders when the main point of institutional finance is in differentiating the two[2].

To become applicable for institutional level operations, we need a framework that acknowledges the realities of price formation. It must recognize the fact that price is not always available and the fact that there can be two or more prices. Theory of price dynamics must go beyond a mere analogy with random walk and be based on nature of involved processes, objective observation and measurement. Development of such theory the subject of this article.

## 2. Measurement in finance

The two assumptions discussed in the previous section have in fact the same and deeper source: they both rely on traders' ability to quickly agree on price and make a transaction. What they miss is a realization that <u>every transaction in financial markets is in fact an elementary act of price measurement</u>. To see this, let's appeal to the definition of market price as it is commonly set in IFRS (International Financial Reporting Standards) and ISDA (International Swaps and Derivatives Association) documents. An asset has a certain

[2] In other words, stochastic framework does provide solutions, but those solutions only work for "spherical chicken in vacuum"

market price when there are parties transacting the asset at that price [7, 8]. From physics perspective a financial transaction is therefore a measurement act that puts a price tag on security.

| Ask size | Price | Bid size |
|---|---|---|
| 100 | 28.20 | |
| 900 | 28.15 | |
| 100 | 27.95 | |
| 300 | 27.90 | |
| 100 | 27.87 | |
| | 27.83 | 100 |
| | 27.82 | 400 |
| | 27.80 | 100 |
| | 27.79 | 1200 |
| | 27.78 | 100 |
| | 27.75 | 300 |

Fig. 2. Typical order book. Red numbers represent orders to sell, green numbers represent orders to buy.

Let's walk through the process of performing price "measurement". Before we begin, we need to define some notions. When trading orders are submitted to the exchange, they are lined up to form the order book, Fig. 2. Buy order with the highest price $s_{bid}$ is called the best bid[3]. Sell order with the lowest price $s_{ask}$ is called the best ask. The difference between the two prices is called the bid-ask spread

$$\Delta = s_{ask} - s_{bid}$$

When buy and sell orders match, they are taken off the book and a transaction is recorded. Price formation occurs as a result of order interaction in the book.

To perform price measurement, we can submit a buy limit order for 100 shares at a test price to see if it's viable[4], Fig. 3a. If the order does not fill, we'll have to adjust the price upwards, Fig. 3b. We will keep doing it until the order fills. At that point measurement is complete and the transaction price is taken as its result, in this case $27.83.

[3] Technically any level in order book can be made of multiple orders lined up as they arrive. Order book can also be built from books on multiple exchanges. However, these details are unessential as we try to keep ideas clear.

[4] For clarity we use an idealized description. Actual process may be much more complex or volatile, but its nature stays the same.

| Ask size | Price | Bid size |
|---|---|---|
| 100 | 28.20 | |
| 1000 | 28.15 | |
| 100 | 27.95 | |
| 300 | 27.90 | |
| 100 | 27.87 | |
| | 27.83 | 100 |
| | 27.82 | 400 |
| | 27.80 | 200 |
| | 27.79 | 1200 |
| | 27.78 | 100 |
| | 27.75 | 300 |

(a)

| Ask size | Price | Bid size |
|---|---|---|
| 100 | 28.20 | |
| 1000 | 28.15 | |
| 100 | 27.95 | |
| 300 | 27.90 | |
| 100 | 27.87 | |
| | 27.83 | 200 |
| | 27.82 | 400 |
| | 27.80 | 100 |
| | 27.79 | 1200 |
| | 27.78 | 100 |
| | 27.75 | 300 |

(b)

Fig. 3. Price measurement process coming from the buying side: (a) Buy order for 100 shares is initially submitted at $27.80 level (size adds to another already existing order). (b) To fill the order its price has to be moved up. Measurement result reads $27.83.

Instead of buying, we could try selling, Fig. 4a. In this case, we will have to keep lowering the price until order fills, which can happen at $27.87, Fig. 4b. The two measurements coming from the opposite sides produce different results, separated by the amount of spread. We can say that price has an uncertainty and spread measures the degree of that uncertainty. Being localized within the spread price remains uncertain until the next transaction.

Of course, rather than using limit orders and gradually adjusting price, we could use market orders for immediate execution. Then the two measurement results would switch places, but it would still be the same two numbers and the same price uncertainty. This uncertainty is an intrinsic property of price and cannot be fully eliminated. It can be diminished by adding more liquidity to the market, but it cannot be fully taken away [9, 10].

For the experiment above we used 100 share orders. Larger orders will produce a yet different result. For example, a market sell order for 700 shares would penetrate three bid levels and stop at the fourth with price $27.79. And a sell order for 700 shares at a limit price above $27.83 (but not very far up) would result in redistribution of the bid orders before any execution happens. Buyers, seeing a large quantity posted for sale, may shift their prices downwards hoping to get a better price from the large seller. Even though additional execution features, such as iceberging, hiding or chaining may reduce the impact, it will still be there, since the order's presence will alter the behavior of other traders over time.

We see that price measurement does not look like a typical classical measurement process. Having different results in different measurement acts indicates that price measurement has quantum features. Instead of a single price that randomly changes in time, there is a spectrum of prices, and one of them gets selected in the process of measurement. The end result of the measurement depends on the SPECIFICS of how that measurement is performed!

| Ask size | Price | Bid size |
|---|---|---|
| 100 | 28.20 | |
| 1000 | 28.15 | |
| 100 | 27.95 | |
| 300 | 27.90 | |
| 100 | 27.87 | |
| | 27.83 | 100 |
| | 27.82 | 400 |
| | 27.80 | 200 |
| | 27.79 | 1200 |
| | 27.78 | 100 |
| | 27.75 | 300 |

(a)

| Ask size | Price | Bid size |
|---|---|---|
| 100 | 28.20 | |
| 900 | 28.15 | |
| 100 | 27.95 | |
| 300 | 27.90 | |
| 200 | 27.87 | |
| | 27.83 | 100 |
| | 27.82 | 400 |
| | 27.80 | 200 |
| | 27.79 | 1200 |
| | 27.78 | 100 |
| | 27.75 | 300 |

(b)

Fig. 4. Price measurement process coming from the selling side. The result reads $27.87.

Following these considerations, quantum theory of price formation was proposed and developed by the author in [9-17], treating price formation as a **quantum-chaotic** process rather than **classical-chaotic** process. In this approach securities are governed by the stochastic price operator, whose spectrum defines the prices that the security can attain when it is transacted. Each transaction represents an elementary act of price measurement. The security can exist in a mixed state with different prices until one of the prices is selected in a trade bringing the security to the state with that price.

Relation of quantum-mechanical framework to stochastic approach is shown in [18-19], whose author takes the reverse approach to derive the financial Schrodinger equation from the traditional stochastics. In [20] the Schrodinger equation is used to explore price uncertainty within the spread by quantizing the potential created by buyers and sellers and leading to gradual change from underpriced state to overpriced as opposed to classical theory, where that boundary is sharp. Inclusion of microstructural dependencies leads to quantum formulation in [21] as well, as the authors model price evolution under linear market impact assumption. An excellent overview with findings and motivation for quantum formulation is given in book [22].

The essential difference between classical and quantum descriptions has been extensively addressed in literature [23, 24]. The fact that price formation has features of a quantum process points that it cannot be fully described by statistical, stochastic, game-theoretical, or any classical-based models. Classical models may be applicable in limited cases but will fail if applied outside of their limits. This is why current attempts to understand spread's behavior within the stochastic calculus sooner or later lead to discrepancies and paradoxes. No matter how many extra variables and dimensions are added to the model, the classical stochastic approach will not provide a complete and consistent description of spread and liquidity, simply because spread is neglected in that approach in the first place.

Without loss of generality we will discuss equities here, although the theory can be applied to any asset class. We will also formulate the equations for asset price, although it would be more feasible to formulate them for asset returns. This is done to make ideas easier to comprehend for broader audience. Lastly, we would like to point out again that actual price formation process as it is observed on the trading screen or in trading negotiations may seem to be much more complex than the process described here. We have incorporated the essential elements in this theory, while the modifications will only result in technicalities that can be adapted by users for their specific cases.

### 3. Bid and ask prices as eigenvalues of price operator

Having formulated the background, let us begin by postulating that security prices are governed by the *price operator* $\hat{S}$, whose eigenvalues represent the spectrum of prices, that the security can attain:

$$\hat{S}\psi_n = s_n\psi_n \tag{1}$$

Elements of the eigenvector $\psi_n$ are the probability amplitudes forming the probabilities of attaining the price corresponding to the $n$-th state:

$$p_n = |\psi_n|^2 \tag{2}$$

Obviously, price operator $\hat{S}$ must be Hermitian since price is a real number. Since prices fluctuate in time, the price operator has to be stochastic:

$$\hat{S}(t + \delta t) = \hat{S}(t) + \delta\hat{S}(t) \tag{3}$$

Price operator can be represented by an $N \times N$ matrix corresponding to $N$ price levels. In the simplest case, it is feasible to keep only 2 levels reflecting the fact, that largest interaction occurs between the "best" orders, in the so-called "top of the book". Let us therefore consider a model, in which the security has only two states: one with price equal to bid price, and the other with price equal to ask price:

$$\psi = \begin{pmatrix} \psi_{ask} \\ \psi_{bid} \end{pmatrix} \tag{4}$$

Writing Eq. (1) in matrix form we have:

$$\begin{pmatrix} s_{11} & s_{12} \\ s_{12}^* & s_{22} \end{pmatrix} \begin{pmatrix} \psi_{ask} \\ \psi_{bid} \end{pmatrix} = s_{ask/bid} \begin{pmatrix} \psi_{ask} \\ \psi_{bid} \end{pmatrix} \tag{5}$$

Eigenvalues $s_{ask}$ and $s_{bid}$ are then expressed through matrix elements as

$$s_{ask} = \frac{s_{11} + s_{22}}{2} + \sqrt{\left(\frac{s_{11} - s_{22}}{2}\right)^2 + |s_{12}|^2} \tag{6a}$$

$$s_{bid} = \frac{s_{11} + s_{22}}{2} - \sqrt{\left(\frac{s_{11} - s_{22}}{2}\right)^2 + |s_{12}|^2} \tag{6b}$$

These equations can be rewritten as

$$s_{ask} = s_{mid} + \frac{\Delta}{2} \quad and \quad s_{bid} = s_{mid} - \frac{\Delta}{2} \tag{7}$$

to represent the two prices spaced out by the spread

$$\Delta = \sqrt{(s_{11} - s_{22})^2 + 4|s_{12}|^2} \tag{8}$$

around the mid price

$$s_{mid} = \frac{s_{11} + s_{22}}{2} = \frac{s_{bid} + s_{ask}}{2} \quad (9)$$

Trivial solution with single price $s$ can be produced by the following price operator:

$$\hat{S} = \begin{pmatrix} s & 0 \\ 0 & s \end{pmatrix} \quad (10)$$

**4. Spread and its statistics**

Let us write the fluctuating matrix elements $s_{ik}$ in the following form, where we factor out the common component $s_{trade}(t)\,\sigma dz$:

$$s_{11}(t + \delta t) = s_{trade}(t) + s_{trade}(t)\,\sigma dz + \frac{\xi}{2} \quad (11a)$$

$$s_{22}(t + \delta t) = s_{trade}(t) + s_{trade}(t)\,\sigma dz - \frac{\xi}{2} \quad (11b)$$

$$s_{12}(t + \delta t) = \frac{\kappa}{2} \quad (11c)$$

Here $\sigma$ is the volatility, $dz \sim N(0,1)$, $\xi \sim N(\xi_0, \xi_1)$ and $\kappa \sim N(\kappa_0, \kappa_1)$[5]. In such setup mid-price and spread are given by equations

$$s_{mid}(t + \delta t) = s_{trade}(t) + s_{trade}(t)\,\sigma dz \quad (12a)$$

$$\Delta = \sqrt{\xi^2 + \kappa^2} \quad (12b)$$

Trade price is then chosen between the $s_{bid}$ and $s_{ask}$ according to probability of their occurrence:

$$s_{trade}(t + \delta t) = \begin{bmatrix} s_{bid} \text{ with probability } |\psi_{bid}|^2 \\ s_{ask} \text{ with probability } |\psi_{ask}|^2 \end{bmatrix} \quad (13)$$

This model includes the traditional Wiener process as a trivial case. Indeed, when $\xi = \kappa = 0$ spread equals zero, and

$$s_{trade}(t + \delta t) = s_{trade}(t) + s_{trade}(t)\,\sigma dz \quad (14)$$

[5] Strictly speaking, $\kappa$ can be a random complex number

which is the traditional Wiener process for $s_{trade}$ showing that the correspondence principle is observed.

Breaking down factors contributing to spread, we see that there are two components:

(a) interaction component $\kappa$, which represents the coupling strength and reflects price uncertainty due to probability transfer between the price levels

(b) intrinsic component $\xi$, which can exist even in the absence of level interactions. This component arises due to liquidity limitation and must include fixed factors such as transaction costs and holding premium.

For $\xi_0 = 0$ and $\kappa_0 = 0$, the probability distribution of $\Delta$ is described by equation:

$$P(\Delta) = \frac{1}{\xi_1 \kappa_1} \Delta e^{-a\Delta^2} I_0(b\Delta^2) \tag{15}$$

where

$$a = \frac{1}{4}\left(\frac{1}{\xi_1^2} + \frac{1}{\kappa_1^2}\right) \quad \text{and} \quad b = \frac{1}{4}\left(\frac{1}{\xi_1^2} - \frac{1}{\kappa_1^2}\right)$$

and $I_0(x)$ is the modified Bessel function of the first kind, and which has the following integral representation:

$$I_0(x) = \frac{1}{\pi}\int_0^{\pi} e^{x\cos(\phi)} d\phi$$

Appearance of $\Delta e^{-a\Delta^2}$ factor is typical of quantum chaotic systems described by random matrices of Gaussian orthogonal ensemble [25]. We can say that <u>spread behaves as a quantum chaotic variable</u>.

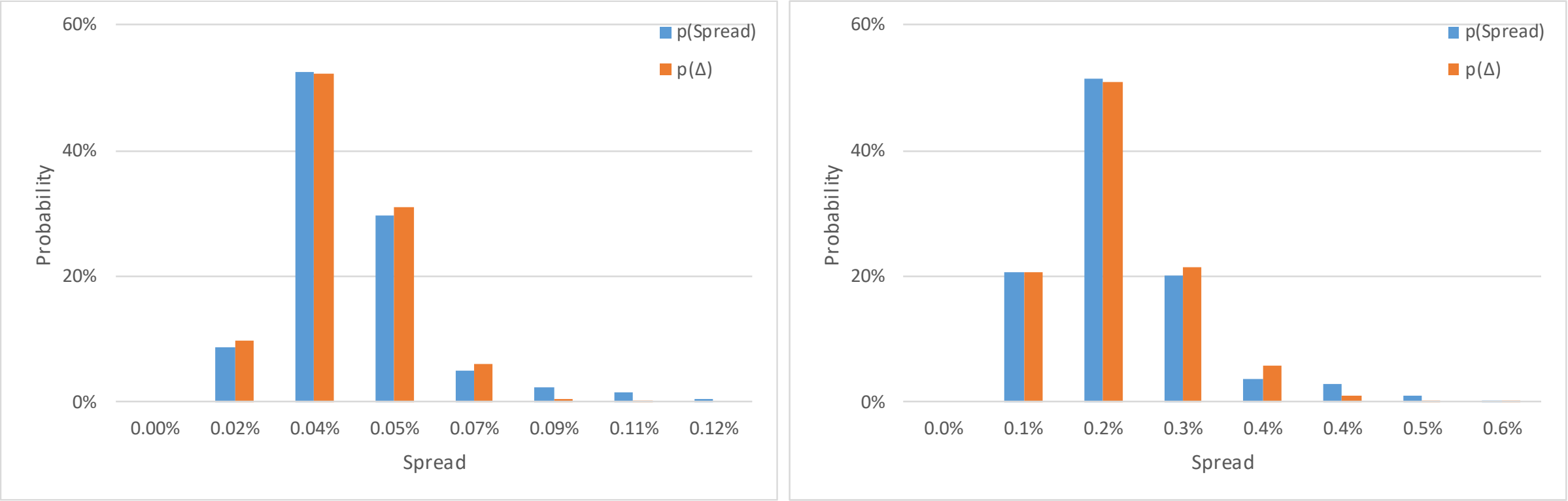

TSLA ALNY

Fig. 5. Bid-ask spread probability distributions and the corresponding fits $p(\Delta)$ for Tesla Inc. (TSLA) and Alnylam Pharmaceuticals, Inc. (ALNY).

Probability distribution of bid-ask spread along with the calibrated model distributions are shown in Fig. 5 for TSLA and ALNY tickers.

**5. State dynamics and execution probability**

Equations Eqs. (11a-11c) allow to model spread statistics but are insufficient for modeling price dynamics. Price dynamics depends not only on the position of bid and ask levels, but also on selection between the two levels, Eqs. (13). To determine it, we need to be able to describe the dynamics of probability amplitude.

Here we must say a few words about derivation of quantum equations in finance. Coming from the probabilistic nature of finance and quantum mechanics some researchers made attempts to simply borrow quantum mechanical equations and draw their analogy with finance. That is incomplete. To do it right, quantum mechanical equations have to be REDERIVED for finance as they apply to the nature of processes in finance while paying attention to assumptions that have been made on the way. This is true even if the resulting equations come out the same. “Borrowing” or “bolting” equations from quantum mechanics onto finance just based on a similarity can lead to distorted picture, which is the opposite to what we are trying to achieve here. Having specified that, let’s follow the standard derivation procedure that can be found in so many textbooks [23, 24] while noting all the assumptions made in the process.

To be able to derive the dynamic equation for $\psi$ we have to rely on superposition principle, which says that a security’s state can always be described as a linear combination of its eigenstates. There is no reasoning allowing to deduce this principle and, in a sense, it must be postulated in finance just the way it was postulated in quantum mechanics with its validity confirmed by subsequent observations [23, 24]. To avoid limitations described in Section 1, the eigenstates must include all possible states, particularly those associated with corporate actions, such as default, M&A, early redemption, delisting, etc. Other special states are those without bidders or without sellers. These states are already included in the framework with $s_{bid} = 0$ and $s_{ask} = \infty$.

Proceeding with the standard derivation procedure, we can say that the derivative $\partial\psi/\partial t$ must come as a result of a linear combination of the components of probability amplitude:

$$\frac{\partial\psi}{\partial t} = \hat{T}\psi, \tag{16}$$

where $\hat{T}$ is some linear operator. If all states $\phi_k$ are taken into account as noted above, this equation must conserve total probability at all times: $p = \sum_k |\phi_k|^2 = 1$. By requiring that $\frac{\partial p}{\partial t} = 0$, it's easy to show that the equation must take form

$$i\tau s_0 \frac{\partial\psi}{\partial t} = \hat{S}\psi \tag{17}$$

where $\tau$ and $s_0$ are some constants[6] with dimensions of time and price, and the operator $\hat{S}$ has to be identified with the price operator to recover the stationary trivial solution with a zero spread and single price:

$$\psi(t) = e^{-i\frac{s\ t}{s_0\tau}}\ \psi(0)$$

Thus, the dynamics of probability amplitudes is governed by Eq. (17). Writing it for a two-level system, we have:

$$i\tau s_0 \frac{d\psi_{ask}}{dt} = s_{11}\psi_{ask} + s_{12}\psi_{bid} \tag{18a}$$

$$i\tau s_0 \frac{d\psi_{bid}}{dt} = s_{12}^{*}\psi_{ask} + s_{22}\psi_{bid} \tag{18b}$$

For time interval in which the matrix elements of the price operator can be considered constant this system of equations has the following solution expressed through model parameters $\xi$ and $\kappa$ [26, 27]:

$$\psi_{ask}(t) = e^{-is_{mid}t}\left\{\left[cos(\phi) - i\frac{\xi}{\Delta}sin(\phi)\right]\psi_{ask}(0) - i\frac{\kappa}{\Delta}sin(\phi)\psi_{bid}(0)\right\} \tag{19a}$$

$$\psi_{bid}(t) = e^{-is_{mid}t}\left\{-i\frac{\kappa}{\Delta}sin(\phi)\psi_{ask}(0) + \left[cos(\phi) + i\frac{\xi}{\Delta}sin(\phi)\right]\psi_{bid}(0)\right\} \tag{19b}$$

where $\phi = \frac{\Delta}{2\tau s_0}t$. Note, that parameter $\sigma$ does not play role in these equations.

[6] Nature of constant $\tau$ is established in [10] to be the characteristic trade time. In this paper it is sufficient to treat it just a time constant.

These equations have to be applied step after step, while updating the coefficients $\xi$ and $\kappa$ at each step. Knowing the execution probabilities now allows to model level selection in Eqs. (13). An example of path generated by the model is shown in Fig. 6.

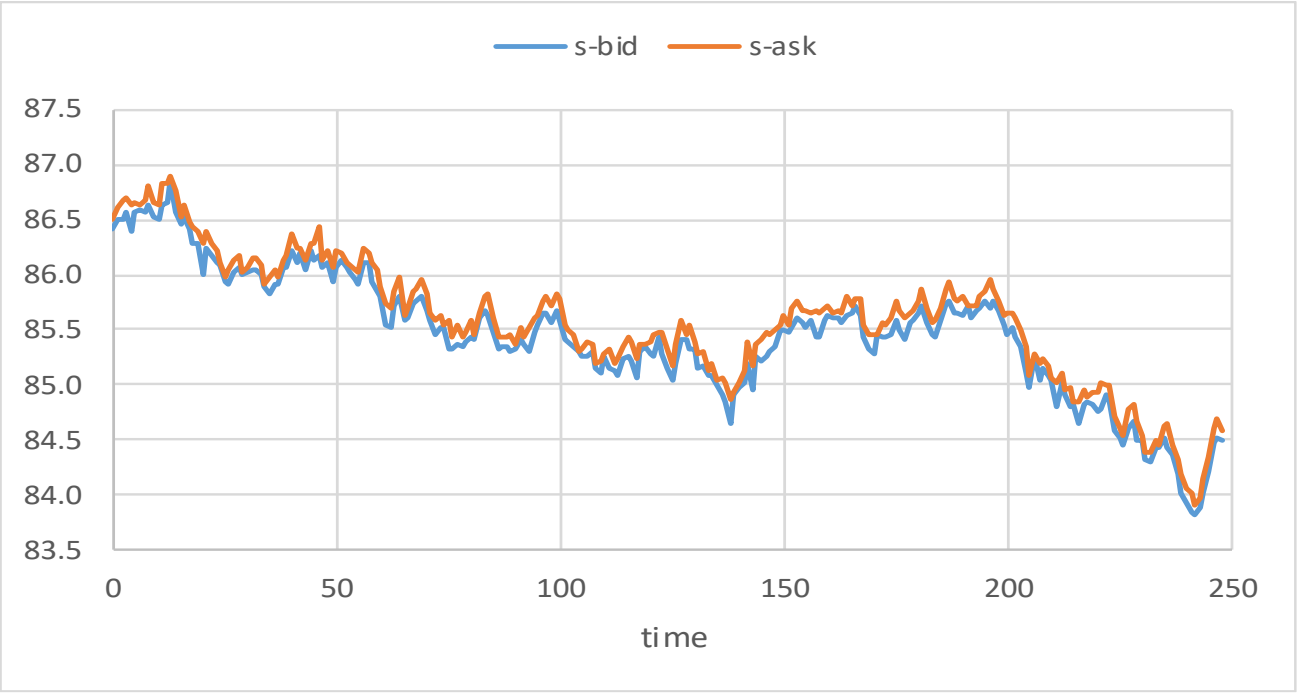

Fig. 6. Typical bid and ask paths in coupled-wave model

Dynamic equations Eqs. (19) are oscillatory in nature. If $\Delta$ were constant, execution probabilities would have oscillated with frequency $\omega = \frac{\Delta}{2\tau s_0}$. However, in practice oscillatory behavior in financial markets is only observed during critical events, such as crashes, price corrections, rallies and rebounds. Apart from these events, absence of interference effects between $\psi_{bid}$ and $\psi_{ask}$ suggests that phase experiences a large random shift at the time of each trade. This may create an impression that phase is unimportant and that the coupled-wave model reduces to just a binomial model [1], in which price after each trade simply takes one of two random values with random probabilities. While resemblance is indeed in place, the binomial model will not produce the relation for spread Eq. (12b), describe spread statistics, or provide any description of its dynamics. A purely stochastic model will have to rely on postulating one or more relations that stem naturally from the quantum framework. Additionally, phase is an integral part of the price process, and we will see below why it becomes important during critical events.

Let us go back to the superposition principle that we used in derivation of Eq. (17). While we did rely on linearity in $\psi$, it does not mean that the model is able to describe only linear effects. Nonlinear effects can come across through coefficients $\xi$ and $\kappa$. Being the components of price operator, these coefficients depend on trading environment and market microstructure. The way they were introduced here, they were assumed random reflecting the fact that the trading environment is generally random. In situations where certain non-random features become prominent, coefficients $\xi$ and $\kappa$ must follow the **material equations** applicable for such situations. These equations may include combinations of $\psi$ and $\psi^*$, thus giving rise to nonlinear effects [28].

Looking at Eqs. (19) something may appear strange about them. For a framework that claims to be universal the fact that probability evolution depends on initial conditions seems unnatural. Surely, the effect of initial conditions can be thought to wash away with time due to large fluctuations. But what determines the initial conditions? How important are they? Does their effect propagate through price evolution? The answer to that lies in the ergodic properties of Eqs. (18), which is discussed in the next section.

## 6. Ergodicity and price formation

To address ergodicity, rather than dealing with probabilities $|\psi_{bid}|^2$ and $|\psi_{ask}|^2$ it is convenient to combine them into a physically meaningful measure – the execution imbalance:

$$I = \frac{|\psi_{ask}|^2 - |\psi_{bid}|^2}{|\psi_{bid}|^2 + |\psi_{ask}|^2} = |\psi_{ask}|^2 - |\psi_{bid}|^2 \tag{20}$$

This quantity can span from 1, when orders are certain to execute at $s_{ask}$ to $-1$, when orders are certain to execute at $s_{bid}$, and is equal to 0 when execution probabilities are equal. **Execution** imbalance should not be confused with **order** imbalance, introduced in [16]. Given that phase is random our interest in system's state is now limited to execution imbalance.

When security trades at its fair value, we must have $|\psi_{bid}|^2 = |\psi_{ask}|^2$ and therefore $I = 0$. This does not mean that price doesn't move, but only that buying and selling probabilities are equal. In reality of course $I$ will fluctuate being symmetrically dispersed around 0 value with some probability distribution $Q(I)$. That probability distribution tends to be narrower for more liquid securities and wider for less liquid ones [16].

Since $I(t)$ depends on initial conditions, it appears that so does $Q(I)$. It is certainly true for a limited observation period. As the system evolves it goes through different microstates of its phase space, and if the system is ergodic, it goes through all microstates, [29-30]. Once it has gone through all and each microstate has made its contribution to $Q(I)$, the initial conditions are no longer singled out, and $Q(I)$ must converge to a definite function, which stays invariant under transformation represented by Eqs. (19).

This is the generally observed picture in balanced market conditions, i.e. when the security trades around its fair value, Fig. 7a. A different picture arises in imbalanced situations, such as price corrections, crashes and rallies. When the number of buyers or sellers is excessive the initial conditions are such that $I$ is close to $-1$ or 1. Their effect in Eqs. (19) persists in time leading to an asymmetric shape of $Q(I)$ and enabling $I$ to stay positive or negative for extended time. The security loses ergodicity and becomes confined to states with $I < 0$ or $I > 0$, even though the other states are still available as shown in Fig. 7b. Such ergodicity loss is accompanied by sharply rising correlations and the volatility-drawdown hysteresis [31, 32], which is the typical ergodicity-breaking behavior.

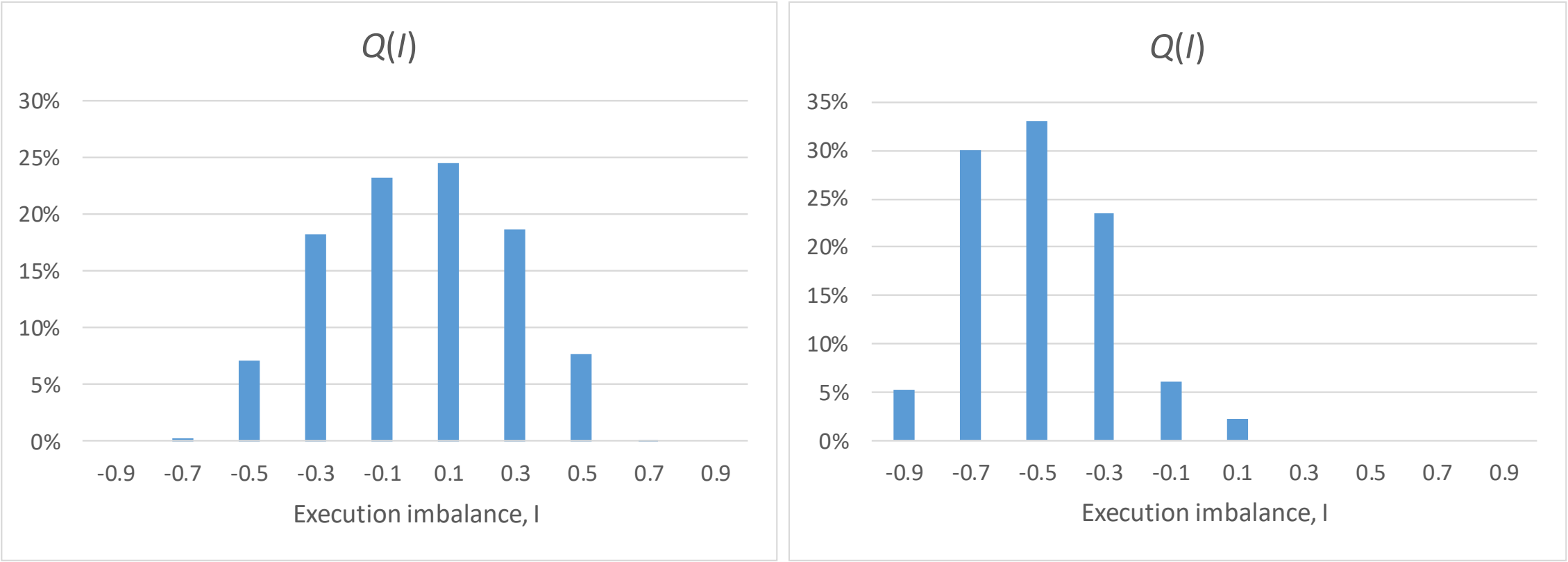

Fig. 7. (a) Probability distribution $Q(I)$ in a balanced market, (b) $Q(I)$ in a crashing market.

This consideration marks a crucial difference of quantum price formation theory with classical economic theory. Classical economics teaches that price forms as a result of balance between supply ($S$) and demand ($D$) at the intersection of supply and demand curves (single number!). When either of them changes, it results in a shift of the intersection point. This results in force, proportionate to the disbalance and acting on price until it reaches its new equilibrium level [34-36]: $\frac{ds}{s} \sim \frac{dD}{D} - \frac{dS}{S}$.

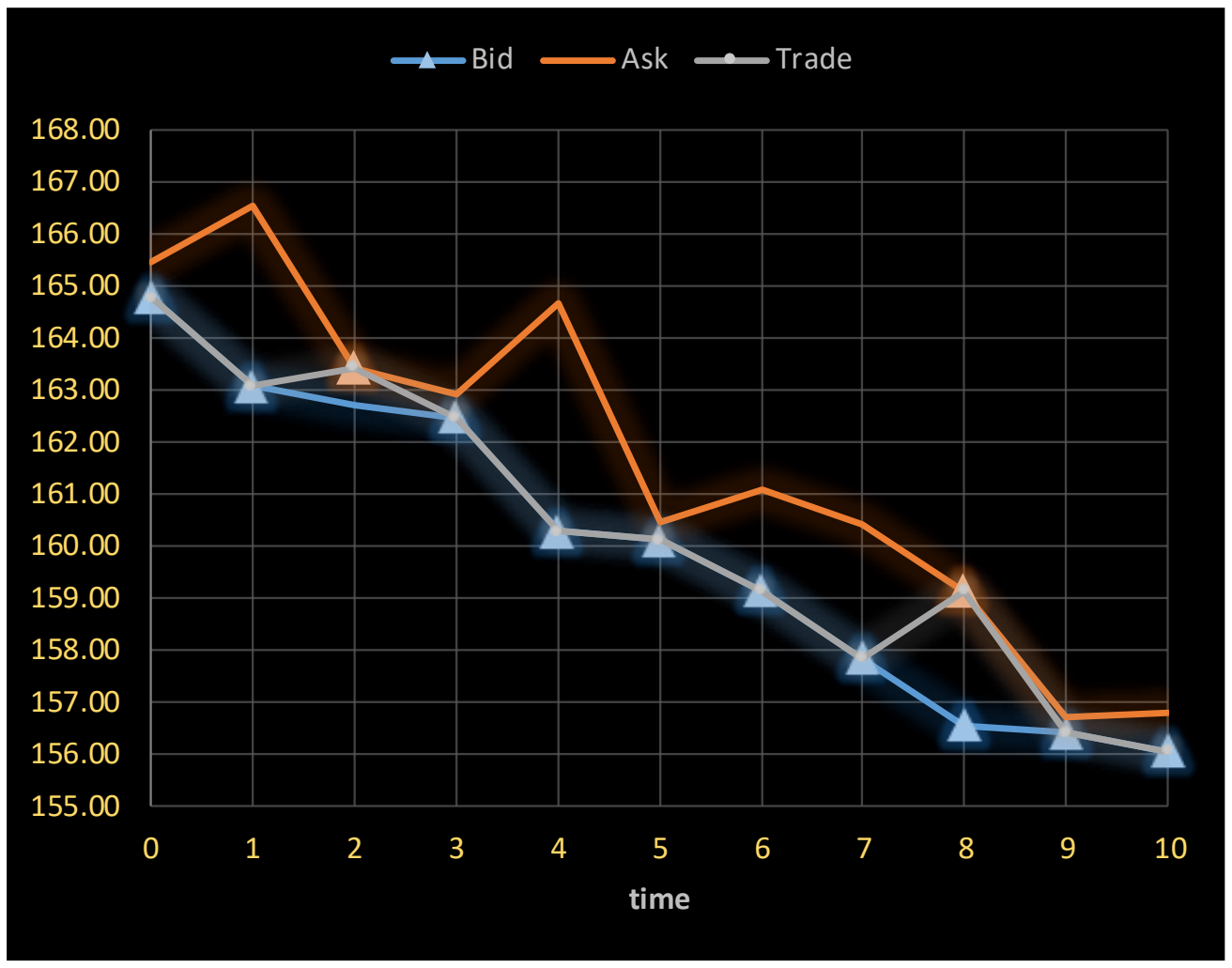

Fig. 8. Directional price move in price crash scenario. Due to ergodicity breaking most executions happen at the bid level, driving the price down without application of any forces.

This is not how it happens in quantum framework. Take a crash scenario as an example. Excessive volume of sellers creates a shift in execution probabilities towards the bid-level. When this happens, the security state shifts towards the bid-level and bid price becomes a more likely outcome of measurement. Once a trade happens at that level, the bid and ask orders will position around the new price level. Once again, the probability shift makes the bid-level a more likely outcome for the next trade. Repeated over many steps, price goes down as shown in Fig. 8. This happens as a result of ergodicity violation in a quantum-chaotic process without any mystical forces acting on price.

Using imbalance, it is now easy to see why the phase comes into play during critical events and is insignificant otherwise. Apart from imbalance $I$ being eliminated at fair value, coupling between levels is also eliminated since there is no reason for probability transfer. This indicates that $\kappa \sim I$ [16]. This dependence is what ensures price stabilization by creating a point of stability with $I = 0$. Taking this into account, a more detailed form of Eq. (12b) can be written:

$$\Delta \approx \sqrt{\xi^2 + a^2 I^2},$$

where $a$ is a proportionality coefficient. For a security at fair value the two terms under the root have the same order of magnitude: $aI \sim \xi$, so that phase evolution is completely random. For an imbalanced security, $aI \gg \xi$ and therefore $\Delta \approx aI + \eta$, where $\eta$ is noise and $I \approx \pm 1$ depending on the direction of the event. This appearance of a persistent term $aI$ is what gives rise to oscillatory patterns in critical events.

## 7. Model extensions

It is important to keep in mind that the coupled-wave model assumes two price levels. It is therefore applicable only if transfer process involves mainly two levels so that the other levels can be neglected. Such assumption can become invalid in situations with high volatility when many levels are penetrated, in the first hour of market opening, or when order size is large. To describe transfer between more levels we need to go back to Eq. (17) and apply it for a larger number of levels, [14]. However, simplifications are available that allow to bypass such complex computations and yet obtain reasonable results. For example, we can combine top $N$ individual levels into multilayer levels with the effective bid and ask prices and apply the coupled-wave model to them:

$$ask_{eff} = \frac{\sum_{i=1}^{N} s_{ask,i}\, n_{ask,i}}{\sum_{i=1}^{N} n_{ask,i}} \quad \text{and} \quad bid_{eff} = \frac{\sum_{i=1}^{N} s_{bid,i}\, n_{bid,i}}{\sum_{i=1}^{N} n_{bid,i}}$$

Here $n_{ask,i}$ and $n_{bid,i}$ are the order sizes on the $i$-th level from top, while $s_{ask,i}$ and $s_{bid,i}$ are the corresponding prices.

It is also possible to model the more readily available high and low levels of the OHLC[7] bar data. In such setup the "high" price would correspond to "ask" level, and the "low" price would correspond to "bid" level. Calibration examples for high-low bars are presented in Fig. 9. However, in this case Eqs. (17) for probability amplitude are inapplicable. Instead, price must execute within high-low interval with a probability distribution corresponding to market data.

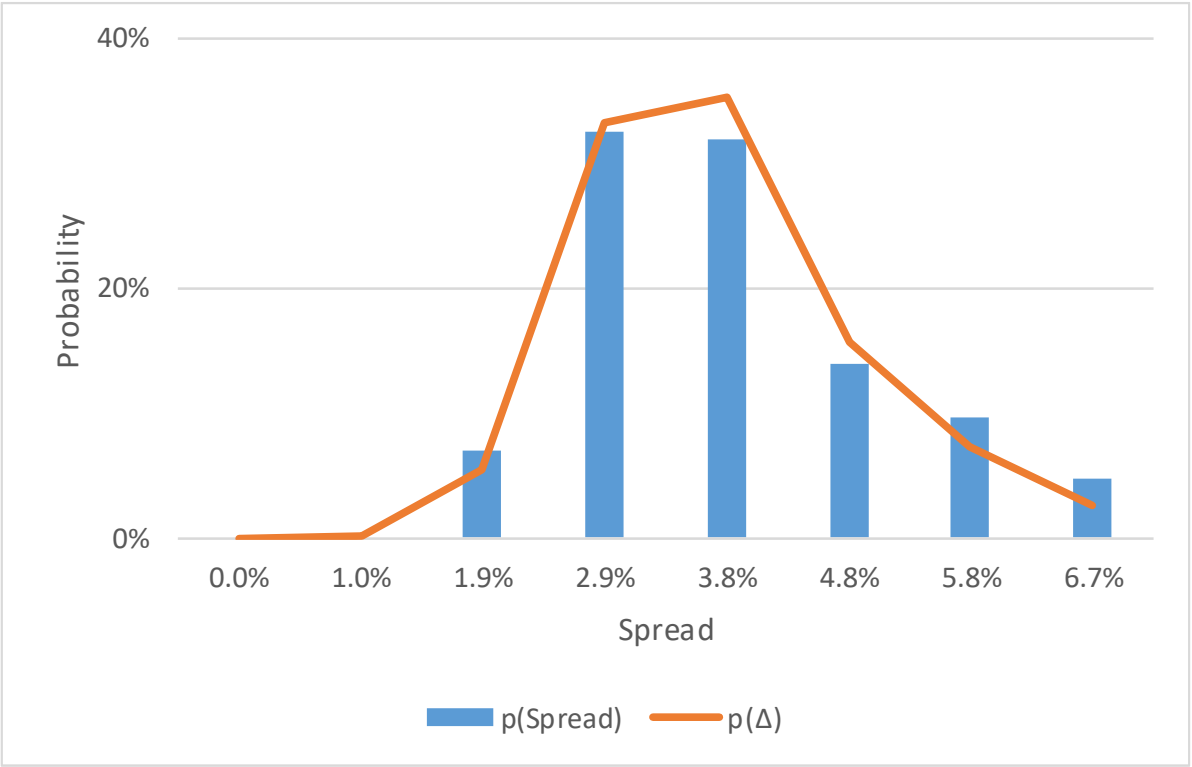

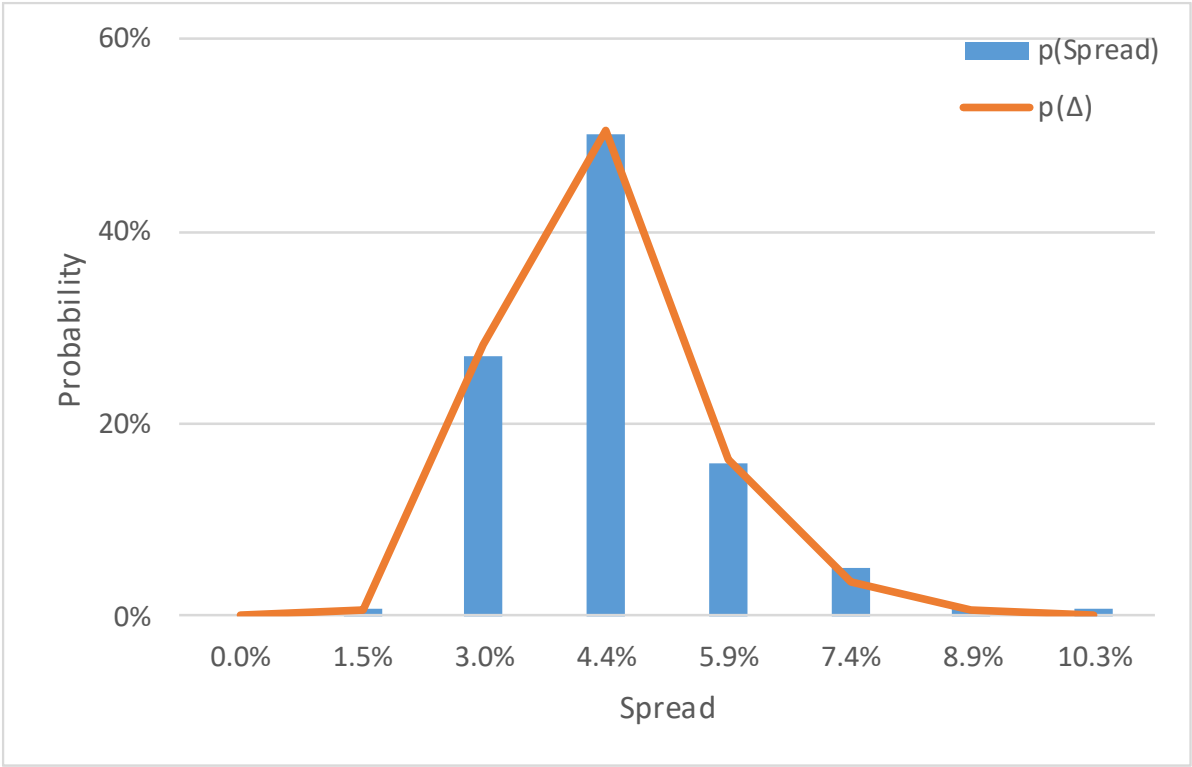

TSLA ALNY

Fig. 9. Probability distribution of daily relative high-low difference and the corresponding fits $p(\Delta)$ for Tesla Inc. (TSLA) and Alnylam Pharmaceuticals, Inc. (ALNY) for Feb 25, 2019 - Sept 20, 2019.

## 8. Discussion

In today's literature we find attempts to apply known solutions of quantum mechanical problems to financial markets. True, as a framework dealing with probabilities such application is too tempting. Attempts to quantize price, volume, draw a similarity with Heisenberg's uncertainty principle, introduction of potential wells, calculation of transition probabilities, etc, many known problems of quantum mechanics try to find their way into finance. This could be valid, if we were sure that equations governing financial processes are identical to the commonly known quantum mechanical equations. Since there is no direct evidence of that, we cannot rely on mere analogy and need to build the theory from scratch without presuming similarities with quantum mechanics. The presented theory is an additional step in that direction.

[7] The traditional open-high-low-close data

Does this mean that price formation has a quantum-mechanical nature? We do not go as far as to claim that there are quants running between trading desks or securities[8]. We tend to think that as a framework quantum mechanical formalism has wider applications than just quantum mechanics and one such area of application is finance. This is not unusual, since many physical systems are described by similar equations [26].

A few remarks about the coupled-wave model are in order. The stochastic form of matrix elements in Eqs. (11a-11c) is chosen to facilitate convenient description of market data for equity asset class. This is not the only possible choice. Format of price operator will vary among asset classes and specific instruments, and analysts can adjust the presented framework to better capture the properties of the instruments they are modeling.

Eigenvalue problem Eq. (1) has been formulated for the price operator and allows negativity of prices. A more appropriate formulation would have been for the logarithmic price or returns, which would exclude possibility of negative prices and include price scaling. In this paper we deliberately based our formulation on price to make ideas easier to grasp.

The model was presented with parameters $\xi$, $\kappa$ and $z$ as random and uncorrelated. While it's a good assumption for a basic model, correlations may exist between all these variables and ultimately, properties of these parameters must come from the material equations. Since our purpose here is to provide the framework, we left details like these out of this article.

Summarizing, we established that security prices behave as quantum-chaotic quantities, not classical-chaotic. This difference comes as a result of the nature of trading, which represents quantum measurement of price in finance. Quantum framework overcomes the default assumptions of stochastic framework about unlimited liquidity and price availability and allows to model processes where execution is a strong factor. The coupled-wave model is the simplest of quantum models that deals with bid and ask prices as different yet connected variables.

[8] Yet, quants do run between trading desks, when "quant" is used as a colloquial reference to "quantitative analysts"

It is important to note that despite numerous similarities with quantum mechanics in our derivation we used no analogy with one, building all blocks from scratch. The similarities come out as a result, not as a presumption.

The need for this framework was generated by the author's own requirements in asset management to be able to price large institutional level orders, execute such orders, execute small orders in HFT format, reliably assess haircuts in REPO transactions, evaluate risk of illiquid positions, and much more. This framework together with all its variations is a universal computational tool allowing to address all these issues. Along with this, the author hopes that it will advance our understanding of financial processes from primitive analogy-based level to the next, physical nature-based level.